\UseRawInputEncoding
\pdfoutput=1
\documentclass[12pt]{article}
\usepackage{tabularx,caption,booktabs}
\usepackage{framed}
\usepackage{doi}
\usepackage[percent]{overpic}
\usepackage{amsfonts,amssymb,graphics,epsfig,verbatim,bm,latexsym,amsmath,url,amsbsy}
\usepackage[affil-it]{authblk}
\usepackage[capposition=top]{floatrow}
\usepackage{float}
\usepackage{graphicx}
\usepackage{booktabs}
\usepackage{multirow}
\usepackage[flushleft]{threeparttable}
\usepackage{enumitem}
\usepackage{mathtools}
\usepackage{qrcode}
\usepackage{titlesec}
\newcolumntype{Y}{>{\centering\arraybackslash}X}
\titleformat{\section}
  {\normalfont\fontfamily{phv}\fontsize{16}{19}\selectfont}{\thesection}{1em}{}
\titleformat{\subsection}
  {\normalfont\fontfamily{phv}\fontsize{14}{17}\selectfont}{\thesubsection}{1em}{}
\titleformat{\subsubsection}
  {\normalfont\fontfamily{phv}\fontsize{14}{17}\selectfont}{\thesubsubsection}{1em}{}

\DeclareMathOperator*{\argmin}{\textsf{arg\,min}}
\DeclareMathOperator*{\ssup}{\textnormal{\textsf{sup}}}

\DeclareMathOperator*{\mmax}{\textnormal{\textsf{max}}}

\newcommand{\cov}{\textsf{cov}}

\newcommand{\Ex}{\textnormal{\textsf{E}}}
\renewcommand{\det}{\textsf{det}}
\renewcommand{\ln}{\textnormal{\textsf{log}}}
\newcommand{\lnsi}{\textnormal{\textsf{log}}^{-1/2}n}
\newcommand{\lni}{\textnormal{\textsf{log}}^{-1}n}
\newcommand{\lns}{\textnormal{\textsf{log}}^{1/2}n}
\newcommand{\ee}{\textnormal{\textsf{e}}}
\newcommand{\eps}{\varepsilon}

\newcommand{\sumt}{\sum_{t\,=\,1}}
\newcommand{\sumi}{\sum_{i\,=\,1}}
\newcommand{\sumj}{\sum_{j\,=\,1}}
\newcommand{\hlambda}{\hat{\lambda}}
\newcommand{\halpha}{\hat{\alpha}}
\newcommand{\st}{\star}
\newcommand{\nn}{\nonumber}
\newcommand{\var}{\textsf{var}}
\newcommand{\No}{\textnormal{\textsf{N}}}

\newcommand{\da}{\dot{a}}

\newcommand{\dda}{\ddot{a}}
\newcommand{\dr}{\dot{r}}
\newcommand{\dl}{\dot{\ell}}
\newcommand{\dg}{\dot{g}}
\newcommand{\df}{\dot{f}}
\newcommand{\ddg}{\ddot{g}}
\newcommand{\ddf}{\ddot{f}}
\newcommand{\mg}{g^{(m)}}
\newcommand{\ma}{a^{(m)}}
\newcommand{\Tr}{\textsf{T}}

\newcommand{\htheta}{\hat{\theta}}

\newcommand{\hgamma}{\hat{\gamma}}
\newcommand{\hkappa}{\hat{\kappa}}

\newtheorem{assumption}{Assumption}
\newtheorem{proposition}{Proposition}

\newtheorem{lemma}{Lemma}

\newtheorem{remark}{Remark}
\newcommand{\btheta}{\bar{\theta}}
\newcommand{\ubtheta}{\munderbar{\theta}}
\newcommand{\bbeta}{\bar{\beta}}
\newcommand{\ubbeta}{\munderbar{\beta}}
\newcommand{\bdelta}{\bar{\delta}}
\newcommand{\ubdelta}{\munderbar{\delta}}
\newcommand{\inn}{\, \in \,}
\newcommand{\eq}{\, = \,}
\newcommand{\lleq}{\, \leq \,}
\newcommand{\ssupt}{\ssup\limits_{\theta \inn [\ubtheta,\btheta]}}
\newcommand{\fra}{\mathfrak{a}}
\usepackage{lineno}

\renewcommand{\thesection}{\arabic{section}}

\renewcommand{\theassumption}{\arabic{section}.\arabic{assumption}}

\renewcommand{\thelemma}{\arabic{section}.\arabic{lemma}}

\usepackage{chngcntr}
\usepackage{apptools}
\usepackage{appendix}
\usepackage{setspace}
\allowdisplaybreaks
\RequirePackage[authoryear]{natbib}
\usepackage[affil-it]{authblk}
\usepackage{accents}
\newcommand\munderbar[1]{%
  \underaccent{\bar}{#1}}

\usepackage{chngcntr}
\counterwithin{remark}{section}

\renewcommand{\thelemma}{\arabic{lemma}}

\usepackage{geometry}
\begin{document}

\newgeometry{margin=1in}
\begin{titlepage}
\thispagestyle{empty}
\title{Two-step estimation in linear regressions with adaptive learning}
\author{{Alexander Mayer}\thanks{Università Ca' Foscari, Venezia; e-mail: \href{mailto:alexandersimon.mayer@unive.it}{alexandersimon.mayer@unive.it}. I would like to thank J\"{o}rg Breitung, Jaqueson Galimberti, Michael Massmann, Sven Otto, and Dominik Wied for helpful discussions. I am grateful to an anonymous referee for valuable comments.}}
	\date{\today}
\maketitle

    \begin{abstract} 
Weak consistency and asymptotic normality of the ordinary least-squares estimator in a linear regression with adaptive learning is derived when the crucial, so-called, `\textit{gain}' parameter is estimated in a first step by nonlinear least squares from an auxiliary model. 
		\end{abstract}

\end{titlepage}
	\restoregeometry


\section{Introduction}

Consider a linear regression with adaptive learning
\begin{equation}\label{eq:ALM}
y_t = \delta_0 + \beta_0a_{t-1}+\eps_t,\;\; t = 1,2,\dots
\end{equation}
where \(\eps_t\) is an error term, \(\delta_0 \in \mathbb{R}\) and the parameter space of \(\beta_0\) is specified below. The variable \(a_{t} \coloneqq a_t(\theta_0)\) is updated according to a stochastic approximation algorithm
\begin{equation}\label{eq:AST}
a_{t}(\theta_0) = a_{t-1}(\theta_0)  + \frac{\theta_0}{t}(y_t - a_{t-1}(\theta_0)), 
\end{equation}
with \(a_0(\theta_0) = \fra_0\) for some initial value \(\fra_0\). The so-called `gain' sequence \(\{\theta_0/t\}_{t\, \geq \,1}\) of positive constants captures the responsiveness to previous prediction mistakes and depends on an unknown gain parameter \(\theta_0\). For \(\theta_0 = 1\), recursion \eqref{eq:AST} reduces to recursive least-squares, which updates \(a_t\) by weighing all observations equally. If, in turn, \(\theta_0 > 1\) (\(\theta_0 < 1\)), then earlier observations receive less (more) weight. The dynamics of \eqref{eq:AST} with this gain specification have been investigated by \cite{masa89} who interpret \(\theta_0\) as a `forgetting factor'; see also \cite{nanu15}, \cite{malmnag16}, or \cite{malmetal21} for recent economic applications. Typically, more general formulations of the type \(y_t =  \beta_0 a_{t-1}x_t + \delta_0x_t  + \eps_t\) are considered within this literature, where \(x_t\) is some stochastic regressor and \(y_{t|t-1}^\ee \coloneqq a_{t-1}x_t\) is viewed as the subjective expectation an economic agent forms about \(y_t\) at time \(t-1\) by estimating the so-called rational expectations equilibrium \(y_t = \alpha_0x_t + \eps_t\), with \(\alpha_0 \coloneqq \delta_0/(1-\beta_0)\); see \cite{evho01}. For brevity, we leave this extension with stochastic \(x_t\) to future research. 

It is well known from the literature on stochastic approximation that the decreasing-gain specification ensures \(a_t \rightarrow \alpha_0\) in an appropriate probabilistic sense; see \cite{bmp90} or \cite{lai03} for an introduction to this literature. Although the resulting asymptotic collinearity makes estimation of the model parameters \(\lambda_0 \coloneqq (\beta_0,\delta_0)^\Tr\) difficult, limiting normality and/or consistency of the ordinary least-squares estimator of \(\lambda_0\) can be established under certain regularity conditions; see \cite{CM18}, \cite{CM19}, and \cite{may21}. The major limitation of these papers is, however, that \(\theta_0\) is treated as a known constant. In what follows, I will thus weaken this assumption and treat the case where the gain parameter \(\theta_0\) is first estimated by nonlinear techniques from a noisy sample of \(\{a_t\}_{t \geq 1}\), an exercise commonly encountered in the empirical macroeconomic learning literature where survey data of the subjective expectations \(y_{t|t-1}^\ee = a_{t-1}x_t\) are used to infer the unknown gain parameter \(\theta_0\); see, e.g., \cite{braev6}, \cite{mapi14}, \cite{malmnag16}, or \cite{bega17}. It is, however, so far unclear which (if any) statistical properties any such estimator possesses. 

The primary objective of this note is thus to derive weak consistency and asymptotic normality of the prominent nonlinear least-squares estimator of \(\theta_0\). This is achieved by combining classical results from  \cite{jenn69} and \cite{lai94} with more recent arguments due to \cite{chawa15} and \cite{wang21}. In addition, the following section also derives the limiting distribution of the feasible ordinary least-squares estimator of \(\lambda_0\) based on the estimated gain and highlights asymptotic equivalence with its unfeasible counterpart under certain parameter combinations, thereby complementing earlier results from \cite{CM18} and \cite{may21} (CMM, henceforth). All proofs are contained in the online appendix.  

\section{Asymptotic Analysis}

\subsection[Nonlinear Least-Squares Estimation of theta]{Nonlinear Least-Squares Estimation of \(\theta_0\)}
Suppose a sample \(\{z_1,\dots,z_n\}\) is observed, where
\begin{equation}\label{noisey}
z_{t} = a_{t-1} + u_t,\;\; t = 1,2,\dots
\end{equation}
\(u_t\) is an error term and the notational convention \(a_t = a_t(\theta_0)\) is recalled. This setting corresponds to the empirical macroeconomic learning literature cited above, where \(\{z_1,\dots,z_n\}\) is interpreted as noisy survey data of \(y_{t|t-1}^\ee = a_{t-1}x_t\) (with the simplifying assumption \(x_t = 1\)), from which the gain parameter is estimated using the least squares method
\[
\htheta_n \coloneqq \argmin\limits_{\theta \inn [\ubtheta,\btheta]} Q_n(\theta),\;\;Q_n(\theta) \coloneqq \sumt^n(z_{t}- a_{t-1}(\theta))^2, \quad 0 < \ubtheta \leq \btheta < \infty.
\]
Here \(a_{t}(\theta)\) denotes the counterpart of Eq. \eqref{eq:AST} that obtains by recursion from Eq. \eqref{eq:AST} upon replacing the true \((\theta_0,\fra_0)^\Tr\) with a candidate value \((\theta,\fra)^\Tr \in [\ubtheta,\btheta] \times \mathbb{R}\).\footnote{It is shown in the appendix that the choice of the initial value is asymptotically inconsequential for a suitably restricted parameter space. Thus, in what follows, the dependence of various quantities on the initial value is suppressed.} 

\subsubsection{Weak Consistency}

The consistency result for \(\htheta_n\) uses a strategy proposed initially by \cite{jenn69}. Define \(D_n(\theta,\theta_0) \coloneqq Q_n(\theta)-Q_n(\theta_0)\). If, for some sequence \(\nu_n \rightarrow \infty\), \(\nu_n^{-1}D_n(\theta,\theta_0) \stackrel{p}{\rightarrow} D(\theta,\theta_0)\) uniformly in \(\theta \in [\ubtheta,\btheta]\), where \(\theta \mapsto D(\theta,\theta_0)\) is continuous and has unique minimum \(\theta_0\) $a.s.$, then \(\htheta_n \stackrel{p}{\rightarrow} \theta_0\). Uniform convergence, in turn, follows from pointwise convergence and a stochastic Lipschitz condition for \(\theta \mapsto Q_n(\theta)\); see, e.g., \citet[Lemma 1]{and92}. In order to establish weak consistency of \(\htheta_n\) along these lines, the following assumptions are imposed.

\renewcommand{\theassumption}{A}
\begin{assumption}\label{ass:B} \(\theta_0 \in (\ubtheta,\btheta)\) so that \textnormal{($a$)} \( 1< \ubtheta < \btheta < \infty\) and \textnormal{($b$)} \(\theta_0(1-\beta_0) > 1/2\). \end{assumption}

\renewcommand{\theassumption}{B}
\begin{assumption}\label{ass:C}\textcolor[rgb]{1,1,1}{.}
\begin{enumerate}
\item[\textnormal{($a$)}] \(\{\eps_t,\mathcal{F}_t\}_{t \geq 1}\), \(\mathcal{F}_t \coloneqq \sigma(\{\eps_j\}_{j \lleq t})\), forms a stationary martingale difference sequence such that \(\Ex[\eps_t^4 \mid \mathcal{F}_{t-1}] < \infty\) a.s. and set \(\sigma_\eps^2 \coloneqq \Ex[\eps_t^2 \mid \mathcal{F}_{t-1}]\). 
\item[\textnormal{($b$)}]\(\{u_t,\mathcal{I}_t\}_{t \geq 1}\), \(\mathcal{I}_t \coloneqq \sigma(\{u_j, a_{j}\}_{j \lleq t})\), forms a stationary martingale difference sequence such that \(\Ex[u_t^4 \mid \mathcal{I}_{t-1}] < \infty\) a.s. and set \(\sigma_u^2 \coloneqq \Ex[u_t^2 \mid \mathcal{I}_{t-1}]\).
\item[\textnormal{($c$)}] The initial value \(\mathfrak{a}_0\) is independently distributed of \(\{u_t\}_{t \geq 1}\) and \(\{\eps_t\}_{t \geq 1}\) such that \(\Ex[\mathfrak{a}_0^4] < \infty.\) 
\end{enumerate}
\end{assumption}
 
\begin{remark}\normalfont
Part \textnormal{($a$)} of Assumption \textnormal{\ref{ass:B}} implies that recent observations are weighed more heavily, a scenario of practical importance as illustrated by, among others, \textnormal{\cite{nanu15}}, \textnormal{\cite{malmnag16}}, or \textnormal{\cite{malmetal21}}. Part \textnormal{($b$)} of Assumption \textnormal{\ref{ass:B}} is not uncommon for the analysis of decreasing gain stochastic approximation; see, e.g., \textnormal{\citet[Theorem 7.10]{evho01}} and \textnormal{\citet[Theorem 13]{bmp90}}. Importantly, \textnormal{($b$)} is equivalent to \(\beta_0 < 1-1/(2\theta_0)\) thereby implying that \(\beta_0 < 1\). More specifically, Assumption \textnormal{\ref{ass:B}}  translates to a short-memory condition for \(y_t\); see \textnormal{\cite{chema17}} and \textnormal{\citet[Corollary 1]{may21}}.  Finally, Assumption \textnormal{\ref{ass:C}} imposes convenient regularity conditions.
\end{remark}	

As a first result, weak consistency is obtained under these assumptions:

	\begin{proposition}\label{prop:consistency} Let \(\nu_n = \ln\,n\) and suppose Assumptions \textnormal{\ref{ass:B}} and \textnormal{\ref{ass:C}} are satisfied.
	\begin{enumerate}
	\item[\textnormal{a)}] For each \(\theta \in [\ubtheta,\btheta]\),  \(\nu_n^{-1}D_n(\theta,\theta_0) \stackrel{p}{\rightarrow} D(\theta,\theta_0) \coloneqq (\theta-\theta_0)^2 d(\theta,\theta_0)\), where
	\[\normalfont
  d(\theta,\theta_0)\coloneqq \left(\frac{\sigma_a}{\theta_0}\right)^2\frac{(2(1-\beta_0)\theta_0-1)\theta-((1-\beta_0)\theta_0-1)}{(2\theta-1)(\theta_0(1-\beta_0)+\theta-1)},
	\]
	with \(\normalfont\sigma_a^2 \coloneqq \sigma_\eps^2\theta_0^2/(2(1-\beta_0)\theta_0-1)\); \(D(\theta,\theta_0)\) is continuous and uniquely minimized at \(\theta = \theta_0\).
	\item[\textnormal{b)}] For each \(\theta_1,\theta_2 \in [\ubtheta,\btheta]\), \(|D_n(\theta_1,\theta_0)-D_n(\theta_2,\theta_0)| \leq h(|\theta_1-\theta_2|)L_n\) for some nonrandom function \(h(x)\) such that \(h(x) \rightarrow 0\) as \(x \rightarrow 0\) and where \(\nu_n^{-1}L_n = O_p(1)\).
	\end{enumerate}
	Consequently, \(\htheta_n \stackrel{p}{\rightarrow} \theta_0\).
	\end{proposition}
	
\begin{remark}\normalfont
The identifying criterion \(D(\theta,\theta_0)\) is non-negative due to the restrictions imposed by Assumption \textnormal{\ref{ass:B}}. The factor \(\sigma_a^2\) coincides with the limiting variance of \(n^{1/2}(a_n-\alpha_0)\), \(\alpha_0 \coloneqq \delta_0/(1-\beta_0)\); see, e.g., \textnormal{CMM}. Using a recent result from \textnormal{\citet[Theorem 2.3]{wang21}} in conjunction with arguments from \textnormal{\cite{lai94}} and \textnormal{CMM}, the stochastic Lipschitz condition is shown to be satisfied for
\[\normalfont
L_n = \ssup\limits_{\theta \inn [\ubtheta,\btheta]}\mmax\left\{\left\vert\sumt^n \da_{t-1}(\theta)u_t\right\vert,\,\sumt^n \da_{t-1}^2(\theta)\right\},\;\; \da_t(\theta) \coloneqq \frac{\textsf{d}\,a_t(\theta)}{\textsf{d}\,\theta}.
\]
\end{remark}	

\subsubsection{Limiting Normality}

In order to derive the limiting distribution, the framework of \cite{chawa15} is adopted. They show that if
\begin{equation}\nn
\begin{split}
\text{($i$)}\;\;&\ssup\limits_{\theta \inn \Theta_n} \ln^{-1}n\bigg\vert\sumt^n\da_t^2(\theta)-\da_t^2(\theta_0)\bigg\vert = o_p(1), \\
\text{($ii$)}\;\;&\ssup\limits_{\theta \inn \Theta_n} \ln^{-1}n\bigg\vert\sumt^n\dda_t(\theta)(a_t(\theta)-a_t(\theta_0))\bigg\vert = o_p(1),\;\; \dda_t(\theta) \coloneqq \frac{\textsf{d}^2\,a_t(\theta)}{\textsf{d}\,\theta^2},
\end{split}
\end{equation}
and
\begin{equation}\nn
\text{($iii$)}\;\;\ssup\limits_{\theta \inn \Theta_n} \ln^{-1}n\bigg\vert\sumt^n\dda_t(\theta)u_t\bigg\vert = o_p(1) 
\end{equation}
holds on \(\Theta_n \coloneqq \{\theta \in [\ubtheta,\btheta]: \ln^{1/2}n|\theta-\theta_0| \leq k_n\}\) for some sequence \(k_n \rightarrow \infty\) as \(n \rightarrow \infty\), then \(\lns(\htheta_n-\theta_0) = Z_n+o_p(1)\),
with
\[
Z_n \coloneqq \lns\left[\sumt^n \da_t^2(\theta_0)\right]^{-1}\sumt^n \da_t(\theta_0)u_t.
\]
The latter quantity, in turn, can be shown to converge weakly to a Gaussian random variable. Proposition \ref{prop:normality} summarizes this argument:

\begin{proposition}\label{prop:normality} If the conditions of Proposition \textnormal{\ref{prop:consistency}} are satisfied, then \(\normalfont\ln^{1/2}n(\htheta_n-\theta_0) = Z_n + o_p(1)\), with \( \normalfont Z_n \stackrel{d}{\rightarrow}\No(0,(\sigma_u/\sigma_{\da})^2)\), where \(\sigma_{\da}^2 \coloneqq d(\theta_0,\theta_0)\) and \(d(\theta,\theta_0)\) has been defined in Proposition \textnormal{\ref{prop:consistency}} so that
\[\normalfont
\sigma_{\da}^2 = d(\theta_0,\theta_0) = \left(\frac{\sigma_a}{\theta_0}\right)^2 \frac{\theta_0(2-\beta_0)-2\theta_0^2(1-\beta_0)-1}{(2\theta_0-1)(1-\theta_0(2-\beta_0))}.
\]
\end{proposition}

\begin{remark}\normalfont  The limit distribution follows from a \textnormal{CLT} due to \textnormal{\cite{dav93}} that allows for degenerate variances. Note that \(\sigma_{\da} > 0\) is ensured by Assumption \textnormal{\ref{ass:B}}. \end{remark}

%

\subsection[Two-Step Estimation of lambda]{Two-Step Estimation of \(\lambda_0\)}
Since \eqref{eq:ALM} is linear in \(\lambda_0\), it is reasonable to estimate \(\lambda_0\) based on the least-squares estimator \(\hlambda_n(\htheta_n)\), where \(\hlambda_n(\theta) \coloneqq \left[\sumt^nw_t(\theta)w_t(\theta)^\Tr\right]^{-1}\sumt^n w_t(\theta)y_t\), with \(w_t(\theta) \coloneqq (1, a_{t-1}(\theta))^\Tr\).
To gain some intuition for the following result, it is instructive 
to note that a second-order Taylor series expansion yields
\begin{equation}\label{eq:ols_taylor1}
\lns(\hlambda_n(\htheta_n)-\lambda_0) = \frac{1}{\sigma_a^{2}}(1,-\alpha_0)^\Tr\left[S_n(\theta_0) + \beta_0Z_n \frac{\dot{S}_n(\theta_0)}{\lns}\right] + o_p(1), 
\end{equation}
with \(Z_n\) from Proposition \ref{prop:normality}, \(S_n(\theta) \coloneqq \lnsi\sumt^n (a_{t-1}(\theta)-\alpha_0)(\eps_t+\beta_0(a_{t-1}(\theta)-a_{t-1}))\), and  \(\dot{S}_n(\theta) = \textsf{d}S_n(\theta)/(\textsf{d}\theta)\). Since CMM have shown that \(S_n(\theta_0)\) is asymptotically normal, the following limiting results are mainly due to Proposition \ref{prop:normality} and the fact that the first-order term of the Taylor series expansion obeys
\begin{equation}\label{eq:ols_taylor2}
\frac{\dot{S}_n(\theta_0)}{\lns} =   \frac{1}{\ln\,n}\sumt^n \da_{t-1}(\theta_0)a_{t-1} + o_p(1) \stackrel{p}{\rightarrow} \sigma_{a\da} \coloneqq \left(\frac{\sigma_a}{\theta_0}\right)^2 \frac{\theta_0(1-(1-\beta_0)\theta_0)}{1-\theta_0(2-\beta_0)}.
\end{equation}
Proposition \ref{cor:lambda} formalizes this argument:

\begin{proposition}\label{cor:lambda} Suppose the conditions of Proposition \textnormal{\ref{prop:consistency}} are satisfied. If, in addition, \(\{u_t\}_{t \geq 1}\) and \(\{\eps_t\}_{t \geq 1}\) are independent, then
\[\normalfont
\ln^{1/2}n(\hlambda_n(\htheta_n)-\lambda_0) \stackrel{d}{\rightarrow}  \No(0_2,(1+B(\theta_0,\beta_0))V_0), \;\; V_0 \coloneqq \frac{2(1-\beta_0)\theta_0-1}{\theta_0^2} \begin{bmatrix} \alpha_0^2 & -\alpha_0 \\ -\alpha_0 & 1 \end{bmatrix},
\]
where \(\normalfont0_2 \coloneqq (0,0)^\Tr\) and
\[
B(\theta_0,\beta_0) \coloneqq \frac{\beta_0^2(2\theta_0-1)((1-\beta_0)\theta_0-1)^2}{((2-\beta_0)\theta_0-1)(1 + \theta_0((1-\beta_0)(2\theta_0-1)-1 ))}\left(\frac{\sigma_u}{\sigma_\eps}\right)^2.
\]
\end{proposition}

\begin{remark}\normalfont The same logarithmic convergence rate has been found by \textnormal{CMM} for the infeasible ordinary least-squares estimator treating the gain as known. The factor \(B(\theta_0,\beta_0) \geq 0\) is due to the presence of the generated covariate. This generated regressor issue vanishes if either \textnormal{(\textit{a})} \(\beta_0 = 0\) or  \textnormal{(\textit{b})} \((1-\beta_0)\theta_0 = 1 \Leftrightarrow \beta_0 = 1-\theta_0^{-1}\), because in both cases \(B(\theta_0,\beta_0) = 0\); i.e., the singular limiting distribution coincides with that of the infeasible estimator \(\hlambda_n(\theta_0)\); see \textnormal{CMM}. The asymptotic equivalence is well known in linear regression if the coefficient on the generated covariate is zero; see, e.g., \textnormal{\cite{pag84}}. Condition \textnormal{(\textit{b})} \((1-\beta_0)\theta_0 = 1\) ensures, in an appropriate probabilistic sense, orthogonality between \(a_t\) and \(\da_t(\theta_0)\). Thus, in view of Eqs. \textnormal{\eqref{eq:ols_taylor1}} and  \textnormal{\eqref{eq:ols_taylor2}}, conditions \textnormal{(\textit{a})} and \textnormal{(\textit{b})} have the same effect on the limiting distribution in that the first-order term in Eq. \textnormal{\eqref{eq:ols_taylor1}} vanishes if they hold true. Under both conditions, the limiting distributions of \(\lns(\htheta_n-\theta_0)\), \(n^{1/2}(a_{n}-\alpha_0)\), and \(\lns(\hlambda_n(\theta_0)-\lambda_0)\) do not depend \textnormal{(\textit{directly})} on \(\beta_0\), the coefficient on \(a_{t-1}\) in model \eqref{eq:ALM}. Finally, independence between \(\{u_t\}_{t \geq 1}\) and \(\{\eps_t\}_{t \geq 1}\), although not essential for deriving Proposition \textnormal{\ref{prop:normality}}, greatly simplifies notation with little loss of insight.
\end{remark}

\subsection[Joint Estimation]{Joint Estimation}
Let me conclude by briefly commenting on two alternative estimation procedures that do not require an auxiliary model based on a noisy sample like Eq. \eqref{noisey}. More specifically, instead of estimating \(\lambda_0 = (\delta_0,\beta_0)^\Tr\) and \(\theta_0\) \textit{sequentially}, one could also try to estimate \(\gamma_0 \coloneqq (\delta_0,\beta_0,\theta_0)^\Tr\) \textit{jointly} based directly on Eq. \eqref{eq:ALM}; see, e.g., \cite{mi07}, \cite{chemama10}, or \cite{bega17b} for empirical applications of joint estimation procedures. In the current context, this would mean to use the following nonlinear least-squares estimator
\[
\hgamma_n \coloneqq \argmin_{\gamma \inn \Gamma} Q_{1,n}(\gamma),\;\; Q_{1,n}(\gamma) \coloneqq \sumt^n(y_t-r_t(\gamma))^2,\;\; r_t(\gamma) \coloneqq \delta+\beta a_{t-1}(\gamma),
\]
with \(\Gamma \coloneqq [\ubdelta,\bdelta] \times [\ubbeta,\bbeta] \times [\ubtheta,\btheta]\), where \(-\infty < \ubdelta < \bdelta < \infty\) and \(-\infty < \ubbeta < \bbeta < 1\). This joint estimator fails, however, important sufficient conditions for nonlinear regression. First, there does not exist a sequence \(\{\tau_n\}_{n \geq 1}\), with \(\tau_n \rightarrow \infty\), such that the sample second moment matrix \(M_n(\gamma) \coloneqq \sumt^n\dr_t(\gamma)\dr_t(\gamma)^\Tr\) of the \(3\times 1\) gradient vector \(\dr_t(\gamma) = (1,a_{t-1}(\theta),\beta \da_{t-1}(\theta))^\Tr\), evaluated at the true parameter values and scaled by \(\tau_n^{-1}\), converges in probability to an invertible matrix, violating, for example,  Eq. (2.4) in \cite{lai94}. Second, \citeauthor{jenn69}'s \citeyearpar{jenn69} condition for weak consistency is violated, as there does not exist a sequence \(\{\nu_n\}_{n \geq 1}\), with \(\nu_n \rightarrow \infty\), such that \(D_{1,n}(\gamma,\gamma_0) \coloneqq Q_{1,n}(\gamma)-Q_{1,n}(\gamma_0)\) converges, when scaled by \(\nu_n^{-1}\), uniformly in probability to a deterministic function with unique minimum. More specifically, we have on \(\Lambda \coloneqq \{\gamma \in \Gamma: \delta = \alpha_0(1-\beta)\} \subset \Gamma\)
\[
\ssup\limits_{\gamma \inn \Lambda}|\lni D_{1,n}(\gamma,\gamma_0)-D_1(\kappa,\kappa_0)| = o_p(1),\;\; \kappa \coloneqq (\beta,\theta)^\Tr,
\]
where 
\begin{equation}
\begin{split}
D_1(\kappa,\kappa_0) \coloneqq \,&\frac{1}{2(1-\beta_0)\theta_0-1}\bigg[
  \beta^2\theta_0^2\\
	\,&+\frac{\beta_0^2\theta^2}{2\theta-1}\bigg(2(1-\beta_0)\theta_0-1+\frac{2\theta_0(\beta_0)((2-\beta_0)\theta_0-1)}{(1-\beta_0)\theta_0+\theta-1}\bigg)\\
	\,&-\frac{2\beta\beta_0(\theta\theta_0(\theta_0(2-\beta_0)-1))}{(1-\beta_0)\theta_0+\theta-1}
\bigg]
\end{split}
\end{equation}
is a continuous non-negative function of \(\kappa\) that attains its minimum of zero if either (1) \(\kappa = \kappa_0\) or, irrespective of whether \(\theta \neq \theta_0\), (2) \(\beta = \beta_0 = 0\). The non-uniqueness due to case (2) reflects an identification failure resulting from the non-separability of \(\beta\) and \(\theta\) in Eq. \eqref{eq:ALM}. On the other hand,
\(
\ssup\limits_{\gamma \inn \Gamma}n^{-1}|D_{1,n}(\gamma,\gamma_0)-\tilde{D}_1(\lambda,\lambda_0)|=o_p(1),
\)
where \(\tilde{D}_1(\lambda,\lambda_0) \coloneqq (\delta-\alpha_0(1-\beta))^2\), with  \(\tilde{D}_1(\lambda,\lambda_0)= 0\) uniformly on \(\Lambda\).\footnote{Interestingly, the estimator of \cite{chemama10} is subject to a similar underidentification issue under deacreasing-gain learning; see \citet[Corollary 6]{may21}.} Even though these conditions are not necessary, the question of whether weak consistency of \(\hgamma_n\) or even asymptotic normality can be shown in spite of their violation is beyond the scope of this note.
 
A little more can be said about the following estimator: Again, the starting point is Eq. \eqref{eq:ALM}, which, after a few rearrangements, yields \(y_t^\st = \beta_0 a_{t-1}^\st + \eps_t\), with  \(y_t^\st \coloneqq y_t - \alpha_0\), \(a_t^\st(\theta) \coloneqq a_t(\theta) - \alpha_0\), and \(a_t^\st \coloneqq a_t^\st(\theta_0)\). Treating, for the sake of argument, \(\alpha_0\) as known, we could first estimate \(\kappa_0 = (\beta_0,\theta_0)^\Tr\) using
\begin{equation}\label{Qsim}
\hkappa_n \coloneqq \argmin\limits_{\kappa \inn \Xi} Q_{2,n}(\kappa),\;\; Q_{2,n}(\kappa) \coloneqq \sumt^n(y_t^\st - \ell_t(\kappa))^2,\;\;\ell_t(\kappa) \coloneqq \beta a_{t-1}^\st(\theta),
\end{equation}
where \(\Xi \coloneqq [\ubbeta,\bbeta]\times[\ubtheta,\btheta]\). Contrary to the preceding discussion, the suitably scaled second moment matrix of the \(2 \times 1\) gradient vector \(\dl_t(\kappa) = (a_{t-1}^\st(\theta),\beta\dot{a}_{t-1}(\theta))^\Tr\) is positive definite with probability approaching one; i.e.
\[
\lni \sumt^n \dl_t(\kappa_0)\dl_t(\kappa_0)^\Tr \stackrel{p}{\rightarrow} K \coloneqq \begin{bmatrix} \sigma_a^2 & \beta_0\sigma_{a\da} \\ \beta_0\sigma_{a\da} & \beta_0^2\sigma_{\da}^2,\end{bmatrix},
\]
where \(K\) is positive definite \textit{as long as} \(\beta_0 \neq 0\). Moreover, \citeauthor{jenn69}'s \citeyearpar{jenn69} condition for weak consistency is satisfied if \(\beta_0 \neq 0\) as \(\lni D_{2,n}(\kappa,\kappa_0)\) converges uniformly in probability to \(D_1(\kappa,\kappa_0)\), which, as mentioned previously, is uniquely minimized at \(\kappa_0\) as long as \(\beta_0 \neq 0\). Thus, by ruling out \(\beta_0 = 0\), we obtain from the above and the results contained in the appendix that
\(
\lns(\hkappa_n-\kappa_0) \stackrel{d}{\rightarrow} \mathcal{N}(0_2,V_2),\;\; V_2 \coloneqq \sigma_\eps^2K^{-1}.
\)
Since, by \citet[Proposition 3]{may21}, \(\halpha_n = \alpha_0 + O_p(n^{-1/2})\), replacing \(\alpha_0\) with \(\halpha_n\) in \eqref{Qsim} is asymptotically inconsequential, making \(\hkappa_n\) a fully feasible estimator. Moreover, in view of \(\alpha_0 = \delta_0/(1-\beta_0)\), an estimator of \(\delta_0\) can be given by \(\hkappa_{n,\delta} \coloneqq \halpha_n(1-\hkappa_{n,\beta})\), where \(\hkappa_{n,\beta}\) is the first component of \(\hkappa_n\). Importantly, if \((1-\beta_0)\theta_0 = 1\), then \(M\) is diagonal and the limiting distributions of both the infeasible and the feasible least-squares estimators from the previous section coincide with that of \(\lns((\hkappa_{n,\delta},\hkappa_{n,\beta})^\Tr-\lambda_0)\). Recall that \((1-\beta_0)\theta_0 = 1\) played a crucial role within the context of Proposition \ref{cor:lambda}. Moreover, the limiting variance of the second component of \(\hkappa_n\) associated with \(\theta_0\), \(\hkappa_{n,\theta}\), say, is given by \((2\theta_0-1)(\theta_0(2-\beta_0)-1)^2/(\beta_0\theta_0)^2\),
which, contrary to the limiting variance of \(\htheta_n\) from Proposition \ref{prop:normality}, does not depend on innovation variances.

\section{Conclusion}

This note derives the singular limiting distribution of the two-step least-squares estimator in a linear regression with estimated gain. Although the limiting distribution is contaminated by the presence of a generated covariate, this generated-regressor problem disappears for certain parameter combinations.

 \nolinenumbers

		\clearpage
 \newgeometry{margin=1.3in}
\appendix

\section[A Preliminaries]{Preliminaries}\label{seq:appA}
\setcounter{equation}{0}
\numberwithin{equation}{section}
\noindent\textbf{Working Formula for \(a_n\) when  \(\{\theta_0,\fra_0\}\) are unknown.}
Given the sample \(\{y_1,\dots,y_n\}\), one obtains a counterpart of Eq. \eqref{eq:AST} that holds for any \(\theta\) and any initial value \(\mathfrak{a} \in \mathbb{R}\) from the following recursive representation
\begin{equation} 
a_{n}(\theta,\fra) = a_{n-1}(\theta,\fra)  + \frac{\theta}{n}(y_n - a_{n-1}(\theta,\fra)),\quad a_0(\theta,\fra) \coloneqq \fra,
\end{equation}
 where the notation \(a_{n}(\theta,\fra)\) is used to make the dependence on the initial value \(\fra\) explicit. Thus, Eq. \eqref{eq:AST} is recovered from the preceding display by using the true parameters \(\{\theta_0,\fra_0\}\); i.e., \(a_{n}(\theta_0,\fra_0)\). Next, by adding and subtracting \(\alpha_0 = \delta_0/(1-\beta_0)\), it follows for all \(n \geq 1\) that
\begin{equation}  \label{eq:ASTgamma} 
\begin{split}
a_{n}(\theta,\fra) - \alpha_0 = \,& (a_{n-1}(\theta,\fra)-\alpha_0)\left(1-\frac{\theta}{n}\right)  + \frac{\theta}{n}(y_n - \alpha_0) \\
= \,& (a_{n-2}(\theta,\fra)-\alpha_0)\left(1-\frac{\theta}{n}\right)\left(1-\frac{\theta}{n-1}\right) \\
\,& + \frac{\theta}{n}(y_n - \alpha_0)+\left(1-\frac{\theta}{n}\right)\frac{\theta}{n-1}(y_{n-1} - \alpha_0) \\
\vdots \,& \\
a_n(\theta,\mathfrak{a})  - \alpha_0 = \,& (\mathfrak{a}-\alpha_0)\Phi_{n,1}(\theta) + \sum_{t \eq 1}^n g_{t,n}(\theta)(y_t-\alpha_0)
\end{split}
\end{equation}
for \(g_{t,n}(\theta) \coloneqq g_{t,n}(\theta,0)\), with
\[
g_{t,n}(\theta,\beta) =  \frac{\theta}{t}\Phi_{t,n+1}(\theta,\beta),\;\;\Phi_{t,n+1}(\theta,\beta) \coloneqq \prod_{j \eq t +1}^n\bigg[1-(1-\beta)\frac{\theta}{j}\bigg].
\]
As a notational convention, I set throughout \(\Phi_{n,t+1}(\theta) \coloneqq \Phi_{n,t+1}(\theta,0)\), \(\Phi_{n}(\theta,\beta) \coloneqq \Phi_{n,1}(\theta,\beta)\), and \(\Phi_{n,n+1}(\theta,\beta) \coloneqq 1\), \(\Phi_{0}(\theta,\beta) \coloneqq 1\).  

\noindent\textbf{Working formula for \(a_n\) when  \(\{\theta_0,\fra_0\}\) are known.} 
We can rewrite Eq. \eqref{eq:ALM} as 
\begin{equation}
y_t -\alpha_0 = \beta_0a_{t-1}^\st+\eps_t, \quad a_t^\st \coloneqq a_t-\alpha_0, \quad \alpha_0 = \frac{\delta_0}{1-\beta_0}.
\end{equation}
As shown e.g. in \citet[Appendix A]{may21}, by inserting the expression for \(y_t -\alpha_0\) from the preceding display back into Eq. \eqref{eq:ASTgamma} one arrives at
\begin{equation}
a_n^\st = (\mathfrak{a}_0-\alpha_0)\Phi_{n} + \sum_{t \eq 1}^n g_{t,n}\eps_t, \quad g_{t,n} \coloneqq g_{t,n}(\theta_0,\beta_0) = \frac{\theta_0}{t}\Phi_{n,t+1},
\end{equation}
with \(\Phi_{n,t+1} \coloneqq \Phi_{n,t+1}(\theta_0,\beta_0)\) and \(\Phi_{n} \coloneqq \Phi_{n,1}\).

\noindent\textbf{Effect of the initial value \(\fra\).} As shown below in Lemma \ref{lemma_aux5}, 
\[
\Ex\big[\ssupt\ssup\limits_{\fra\inn\mathbb{R}}|Q_n(\theta,\fra)-Q_n^\dagger(\theta)|\big] = O(1), 
\]
with
\[
Q_n(\theta,\fra) \coloneqq \sumt^n(z_t-a_{t-1}(\theta,\fra))^2,\quad Q_n^\dagger(\theta) \coloneqq \sumt^n(z_t-a_{t-1}^\dagger(\theta))^2,
\]
where, 
\begin{equation}\label{eq:ASTtheta}
a_n^\dagger(\theta) = \begin{cases}
\mathfrak{a}_0 & \text{if } n = 0, \\
\alpha_0 + \sum_{t \eq 1}^n  g_{t,n}(\theta)(y_t-\alpha_0) & \, \text{otherwise}.
\end{cases}
\end{equation}
Therefore, I redefine \(a_n(\theta) = a_n^\dagger(\theta)\) and work in the following with \(a_n(\theta)\) instead of \(a_n(\theta,\fra)\). For future reference, set
\[
a_n^\st(\theta) \coloneqq a_n(\theta)-\alpha_0.
\]

\noindent\textbf{Notational conventions.} 
Next, define \(\normalfont \mg_{t,n}(\theta) \coloneqq \textsf{d}^m\,g_{t,n}(\theta)/(\textsf{d}\,\theta^m)\), and similarly \(\normalfont \ma_{n}(\theta) \coloneqq \textsf{d}^m\,a_{n}(\theta)/(\textsf{d}\,\theta^m)\), \(m \in \mathbb{N}_+\). For the first four derivatives, the notation \(\dg_{t,n}(\theta)\), \(\ddot{g}_{t,n}(\theta)\), \(\dddot{g}_{t,n}(\theta)\), and \(\ddddot{g}_{t,n}(\theta)\) is used; similarly for \(a_{n}(\theta)\). Finally, set \(\dg_{t,n} \coloneqq \dg_{t,n}(\theta_0)\), \(\da_n \coloneqq \da_n(\theta_0)\) and let \(c_0 \coloneqq 1-\beta_0\).


\section[B Proof of Main Results]{Proof of main results}

\subsection{Proof of Proposition \ref{prop:consistency}}
\underline{Part a).} 
Note that \(D_n(\theta,\theta_0) = S_n(\theta,\theta_0)-2V_n(\theta,\theta_0)\), where \(S_n(\theta,\theta_0) \coloneqq \sumt^n(a_{t-1}^\st-a_{t-1}^\st(\theta))^2\) and \(V_n(\theta,\theta_0) \coloneqq \sumt^n(a_{t-1}^\st-a_{t-1}^\st(\theta))u_t\). Begin with \(S_n(\theta,\theta_0)\) and consider
\[
\lni S_n(\theta,\theta_0) =  \lni \sumt^n a_{t-1}^{\st 2}(\theta) + \lni \sumt^n a_{t-1}^{\st 2}   - 2\lni \sumt^n a_{t-1}^\st a_{t-1}^\st(\theta).
\]
By Lemma \ref{lemma_aux3}, \(\lni S_n(\theta,\theta_0) \stackrel{p}{\rightarrow} D(\theta,\theta_0)\) for any \(\theta \in [\ubtheta,\btheta]\). Similarly, it follows readily that \(\lni V_n(\theta,\theta_0) = o_p(1)\) pointwise in \(\theta \in [\ubtheta,\btheta]\). This proves the claim.

\noindent \underline{Part b).} The claim follows from Lemma \ref{lemma_aux2}.

\subsection{Proof of Proposition \ref{prop:normality}}
 First, conditions ($i$), ($ii$), and ($iii$) from \cite{chawa15}, stated in the main text, are subsequently proven. \underline{Condition ($i$).} Let \(k_n \rightarrow \infty\) such that \(k_n = o(\ln^{1/2}n)\). 
\begin{equation}\nn
\begin{split}
\ssup\limits_{\theta \inn \Theta_n} \ln^{-1}n\bigg\vert\sumt^n\da_t^2(\theta)-\da_t^2(\theta_0)\bigg\vert \stackrel{(1)}{\leq} \,& 2(k_n/\ln^{1/2}n)   \ln^{-1}n\ssup\limits_{\theta \inn [\ubtheta,\btheta]}\bigg\vert\sumt^n\dda_t(\theta)\da_t(\theta)\bigg\vert \\
\stackrel{(2)}{\leq} \,& 2(k_n/\ln^{1/2}n) \bigg[\ln^{-1}n\sumt^n\ssup\limits_{\theta \inn [\ubtheta,\btheta]}\dda_t^2(\theta)\bigg]^{1/2}\\
\,& \hspace*{3cm} \times \bigg[\ln^{-1}n\sumt^n\ssup\limits_{\theta \inn [\ubtheta,\btheta]}\da_t^2(\theta)\bigg]^{1/2} \\
\stackrel{(3)}{=} \,& O_p(k_n/\ln^{1/2}n) = o_p(1).
\end{split}
\end{equation}
Explanations: (1) uses the mean-value theorem and the fact that for any \(\theta \in \Theta_n\) one has \(|\theta-\theta_0| \leq  k_n/\lns\); (2) is due to Cauchy-Schwarz; (3) follows from Lemma \ref{lemma_aux3}. \underline{Condition ($ii$).} Follows by similar arguments as condition ($i$). \underline{Condition ($iii$).} Consider a sequence \(\ell_n \rightarrow \infty\) such that \(\Theta_n \subset \cup_{j \eq 1}^N\Theta_{n,j}\), where \(\Theta_{n,j} \coloneqq \{\theta \in [\ubtheta,\btheta]: |\theta-\theta_j| \leq \ell_n^{-1}\}\) and the integer \(N \coloneqq N(n)\) satisfies \(N = O(\ell_nk_n/\ln^{1/2}n)\). Now, by the triangle inequality
\begin{equation}\nn
\begin{split}
\ssup\limits_{\theta \inn \Theta_n} \ln^{-1}n\bigg\vert\sumt^n\dda_t(\theta)u_t\bigg\vert \leq \,&  \mmax\limits_{1 \lleq j \lleq N}\ssup\limits_{\theta \inn \Theta_{n,j}} \ln^{-1}n\bigg\vert\sumt^n(\dda_t(\theta)-\dda_t(\theta_j))u_t \bigg\vert \\
\,& + \lni\mmax\limits_{1 \lleq j \lleq N} \bigg\vert\sumt^n\dda_t(\theta_j)u_t \bigg\vert \\
\stackrel{(1)}{\leq} \,& \ell_n^{-1} \ssup\limits_{\theta \inn [\ubtheta,\btheta]} \ln^{-1}n\bigg\vert\sumt^n \dddot{a}_t(\theta)u_t \bigg\vert \\
\,&+ \lni\mmax\limits_{1 \lleq j \lleq N} \bigg\vert\sumt^n\dda_t(\theta_j)u_t \bigg\vert \\
\stackrel{(2)}{=} \,& O_p(\ell_n^{-1}) + O_p(\sqrt{\ln(N)/\ln\, n}) = o_p(1)
\end{split}
\end{equation}
using that \(\log(N)/\log(n) = o(1)\) for suitable choices of \(\ell_n\) and \(k_n\). The penultimate equality (2) warrants some additional explanation: Begin with the first summand on the right-hand side of inequality (1). Consider the cover \([\ubtheta,\btheta] \subset \cup_{j \eq 1}^M \tilde{\Theta}_{n,j}\), where \(\tilde{\Theta}_{n,j} \coloneqq \{\theta \in [\ubtheta,\btheta]: |\theta -\theta_j| \leq n^{-1}\}\) and the integer \(M \coloneqq M(n)\) satisfies \(M = O(n^{-1})\). Then, using the mean-value theorem and Cauchy-Schwarz, we can bound the first summand on the right-hand side of inequality (1) by \(\ell_n^{-1}\) times
\begin{equation}\nn
\begin{split}
n^{-1/2}  \bigg[\ln^{-1}n \sumt^n \ssup\limits_{\theta \inn [\ubtheta,\btheta]} \ddddot{a}_{t-1}^2(\theta)u_t^2 \bigg]^{1/2} \,&+ \ln^{-1}n  \mmax\limits_{1 \lleq j \lleq M} \bigg\vert\sumt^n\dddot{a}_{t-1}(\theta_j)u_t \bigg\vert\\
\;& = O_p(n^{-1/2}) + O_p(1),
\end{split}
\end{equation}
using \citet[Theorem 2.3]{wang21} and Lemma \ref{lemma_aux2}. Similarly, since
\[
\sumt^n \ssup\limits_{\theta \inn [\ubtheta,\btheta]}\ddot{a}_{t-1}^2(\theta)u_t^2 = O_p(\ln\,n),
\]
\citet[Theorem 2.3]{wang21} yields \(\mmax\limits_{1 \lleq j \lleq N} \bigg\vert\sumt^n\dda_t(\theta_j)u_t \bigg\vert = O_p(\sqrt{\ln(n)\textsf{log}(N)})\). Next, we derive the limiting distribution of \(Z_n = \lns\big[\sumt^n \da_t^2\big]^{-1}\sumt^n \da_tu_t\). By Lemma \ref{lemma_aux3}, \(\lni \sumt^n \da_t^2 = \sigma_{\da}^2 + o_p(1)\). Moreover, Corollary 2.2 from \cite{dav93} in conjunction with Lemma \ref{lemma_aux3} and Lemma \ref{lemma_aux4} yields \(\lnsi \sumt^n \da_tu_t \stackrel{d}{\rightarrow} \textsf{N}(0,\sigma_\eps^2\sigma_{\da}^2)\). This completes the proof.


\subsection{Proof of Proposition \ref{cor:lambda}}
The proof strategy follows the line of argument employed in the proof of \citet[Proposition 1]{may21} pertaining to the case \(\theta = \theta_0\). Consider \(\hlambda_n(\theta)-\lambda_0 = M_n(\theta)^{-1}Y_n(\theta)\),
where \(Y_n(\theta) \coloneqq \sumt^n w_t(\theta)\eps_t(\theta)\), \(M_n(\theta) \coloneqq \sumt^n w_t(\theta)w^\Tr_t(\theta)\), and
\[
\eps_t(\theta) \coloneqq \eps_t+\beta_0(a_{t-1}-a_{t-1}(\theta)) = \eps_t+\beta_0(a^\st_{t-1}-a^\st_{t-1}(\theta)),
\]
recalling, for convenience, that \(a^\st_t(\theta) = a_t(\theta) - \alpha_0\) and \(a_t = a_t(\theta_0)\). Next, use that \(M_n(\theta)^{-1} = Q_n(\theta)/\det(M_n(\theta))\), with
\begin{equation} \label{eq:Q}
Q_n(\theta) \coloneqq \sumt^n\begin{bmatrix}\displaystyle 1 & \displaystyle -a_{t-1}(\theta)  \\ \displaystyle-a_{t-1}(\theta) & \displaystyle a_{t-1}^2(\theta) \end{bmatrix}
\end{equation}
and
\begin{align} \label{eq:dW}
\det\,M_n(\theta)\coloneqq n\sumt^n a_{t-1}^{\st 2}(\theta)-\bigg[\sumt^n a_{t-1}^\st(\theta)\bigg]^2, \quad \theta \in [\ubtheta,\btheta].
\end{align}
First, I show that \(n^{-1}\ln^{-1}(n)\, \det\,M_n(\htheta_n) = \sigma_a^2 + o_p(1)\). Since \(\htheta_n - \theta_0 = O_p(\ln^{-1/2}n)\), it follows from the mean-value theorem and Lemma \ref{lemma_aux2}  that
\begin{equation}\nn
\begin{split}
\frac{1}{\ln\,n} \sumt^n (a_{t-1}^{\st 2}(\htheta_n)-a_{t-1}^{\st 2}) \,& \leq O_p(\ln^{-1/2}n)  \frac{1}{\ln\,n}  \ssupt\bigg\vert \sumt^n a_{t-1}^{\st}(\theta)\da_{t-1}(\theta)\bigg\vert \\
\,& = o_p(1).
\end{split}
\end{equation}
The final stochastic order is due to Cauchy-Schwarz and a similar argument as that used for step (3) in the proof of Condition ($i$) of Proposition \ref{prop:normality}.
Similarly, one obtains \(n^{-1/2} \sumt^n (a_{t-1}^\st(\htheta_n)-a_{t-1}^\st) = o_p(1)\). Moreover, by item ($ii$) of Lemma \ref{lemma_aux1}, we have \(\ln^{-1}n \sumt^n a_{t-1}^{\st 2} = \sigma_a^2 + o_p(1)\) and the claim follows. Next, the same algebraic rearrangements as in the proof of \citet[Proposition 1]{may21} yield
\begin{equation}\nn
\begin{split}
\frac{Q_n(\theta)/n}{\ln^{1/2}n}\sumt^n w_t(\theta)\eps_t(\theta) = \,&e_\alpha  \frac{1}{\ln^{1/2}n} \sumt^na_{t-1}^\st(\theta)\eps_t(\theta)\\
\,& - e_\alpha  \frac{1}{\ln^{1/2}n}\bigg[n^{-1/2} \sumt^n a_{t-1}^{\st}(\theta)\bigg] U_n(\theta) \\
\,& + e_1   \sqrt{(\ln\,n)/n}\bigg[\frac{1}{\ln n}\sumt^na_{t-1}^{\st 2}(\theta)\bigg]   U_n(\theta) \\
\,& - e_1  n^{-1/2}\bigg[\frac{1}{n^{1/2}}\sumt^na_{t-1}^{\st}(\theta)\bigg]\bigg[\frac{1}{\ln^{1/2}n}\sumt^na_{t-1}^\st(\theta)\eps_t(\theta)\bigg]\\
\eqqcolon \,&  e_\alpha A_n(\theta) -\frac{1}{\ln^{1/2}n} e_\alpha B_n(\theta)U_n(\theta) + \frac{1}{\ln^{1/2}n} e_1 C_n(\theta) U_n(\theta) \\
\,&- n^{-1/2} e_1 B_n(\theta) A_n(\theta), \quad \theta \in [\ubtheta,\btheta],
\end{split}
\end{equation}
say, where \(e_\alpha \coloneqq (1,-\alpha_0)^\Tr\), \(e_1 \coloneqq (0,1)^\Tr\), \(U_n(\theta) \coloneqq n^{-1/2} \sumt^n \eps_t(\theta)\), and \(A_n(\theta)\), \(B_n(\theta)\) have been implicitly defined. The idea is to show that \(U_n(\htheta_n) = O_p(1)\), \(B_n(\htheta_n) = O_p(1)\), and \(C_n(\htheta_n) = O_p(1)\). The claim then follows from the weak convergence of \(A_n(\htheta_n)\) to a Gaussian random variable. Begin with \(U_n(\htheta_n)\) and note that \(n^{-1/2}\sumt^n\eps_t = O_p(1)\). Thus, 
\[
U_n(\htheta_n) = O_p(1) + \beta_0n^{-1/2}\sumt^n(a_{t-1}^\st - a_{t-1}^\st(\htheta_n)).
\]
By the mean-value theorem and Cauchy-Schwarz, one gets
\begin{equation}\label{eq:rootn}
\begin{split}
n^{-1/2}\sumt^n(a_{t-1}^\st - a_{t-1}^\st(\htheta_n)) \leq \,&|\htheta_n-\theta_0| \ssupt n^{-1/2}|\sumt^n\da_{t-1}(\theta)|\\
 \leq  \,& T_n \bigg[\frac{1}{\ln\,n}\sumt^n \ssupt \da_{t-1}^2(\theta) \bigg]^{1/2},
\end{split}
\end{equation}
where \(T_n \coloneqq \lns(\htheta_n-\theta_0)\) and the term in brackets is, by Lemma \ref{lemma_aux2}, stochastically bounded. Next, \(B_n(\htheta_n)\) and \(C_n(\htheta_n)\) can be treated similarly using Lemma \ref{lemma_aux2} in conjunction with \citet[Lemma B.5]{may21}.	Now, consider 
\begin{equation}
\begin{split}
A_n(\theta) =  \,& \frac{1}{\ln^{1/2}n} \sumt^na_{t-1}^\st(\theta)\eps_t(\theta) \\
=\,&  \frac{1}{\ln^{1/2}n} \sumt^n a_{t-1}^\st\eps_t \\
\,&  +  \frac{1}{\ln^{1/2}n} \sumt^n(a_{t-1}^\st(\theta)-a_{t-1}^\st)\eps_t\\
\,&+ \beta_0\frac{1}{\ln^{1/2}n} \sumt^n  a_{t-1}^\st(\theta)(a_{t-1}^\st(\theta)-a_{t-1}^\st) \\
\eqqcolon \,&   A_{1,n} + A_{2,n}(\theta)+ \beta_0A_{3,n}(\theta),
\end{split}
\end{equation}
say. A second-order Taylor series expansion around \(\theta_0\) yields
\[
A_{2,n}(\htheta_n) =  T_n Z_{2,n} + \ln^{-1/2}(n) \,T_n^2 R_{2,n}(\theta_n)/2,\;\;T_n \coloneqq  \lns(\htheta_n-\theta_0),
\]
with
\[
Z_{2,n} \coloneqq  \frac{1}{\ln\,n} \sumt^n \da_{t-1}\eps_t,\;\; R_{2,n}(\theta_n) \coloneqq \frac{1}{\ln\,n} \sumt^n \dda_{t-1}(\theta_n)\eps_t,
\]
where \(\theta_n\) lies on the line segment connecting \(\theta_0\) and \(\htheta_n\). Since  \(\Ex[Z_{2,n}] = 0\) and \(\var[Z_{2,n}] = o(1)\), \(A_{2,n}(\htheta_n) = o_p(1)\) follows because, by Lemma \ref{lemma_aux2}, \(\ssup_\theta |R_{2,n}(\theta)| = O_p(1)\).  Next, a second-order Taylor series expansion of \(A_{3,n}(\htheta_n)\) around \(\theta_0\) yields
\(
A_{3,n} =  T_n Z_{3,n} + \ln^{-1/2}(n)\, T_n^2 R_{3,n}(\theta_n)/2,
\)
with \(Z_{3,n} \coloneqq  \ln^{-1}(n) \sumt^n \da_{t-1}a_{t-1}^\st\), and
\[
R_{3,n}(\theta_n) \coloneqq \lni \sumt^n ((2a_{t-1}^\st(\theta_n)-a_{t-1}^\st)\dda_{t-1}(\theta_n)+2\da_{t-1}^2(\theta_n)).
\]
It follows from Lemma \ref{lemma_aux3} ($iv$) that
\[
Z_{3,n} \stackrel{p}{\rightarrow} \sigma_{a\da} \coloneqq \left(\frac{\sigma_a}{\theta_0}\right)^2\frac{\theta_0(1-c_0\theta_0)}{1-\theta_0(1+c_0)},
\]
and \(\ssup_\theta |R_{3,n}(\theta)| = O_p(1)\). Moreover, due to Proposition \ref{prop:normality}, one gets 
\[
T_n = \frac{1}{\sigma^2_{\da}\ln\,n}\sumt^n \da_{t-1}u_t + o_p(1).
\]
Hence, we can write \(A_n(\htheta_n) = \lnsi\sumt^n z_t + o_p(1)\), where
\(z_t \coloneqq a_{t-1}^\st\eps_t+ \pi_0\da_{t-1}u_t,\) with \(\pi \coloneqq (1-c_0)\sigma_{a\da}\sigma_{\da}^{-2}\). Note that \(\{z_t,\mathcal{G}_t\}_{t \geq 1}\), where \(\mathcal{G}_t = \mathcal{I}_t \vee \mathcal{F}_t\), forms a martingale difference sequence. As \(\var[z_t] = \var[a_{t-1}^{\st}]\sigma_\eps^2+\var[\da_{t-1}]\pi^2\sigma_u^2\), one obtains, by the proof of Lemma \ref{lemma_aux3},
\(
t\var[z_t] = \sigma_\eps^2\sigma_a^2+\pi^2\sigma_u^2\sigma_{\da}^2+ o(1).
\)
Thus, upon collecting terms,
\[
\frac{1}{\ln\,n}\sumt^n \var[z_t] =  \sigma_z^2 + o(1)\;\; \sigma_z^2 \coloneqq \sigma_a^4\frac{2c_0\theta_0-1}{\theta_0^2}(1+B(\theta_0,\beta_0)).
\]
Since, by Lemma \ref{lemma_aux4}, \(\Ex[z_t^4] = O(t^2)\), the CLT of \citet[Corollary 2.2]{dav93} yields \(\lnsi \sumt^n z_t \stackrel{d}{\rightarrow} \textsf{N}(0,\sigma_z^2)\). As shown above \(n^{-1}\ln^{-1}(n)\, \det(M_n(\htheta_n)) = \sigma_a^2 + o_p(1)\), which proves the claim.

\section[C Auxiliary Results]{Auxiliary Results}\label{seq:appC0}

\renewcommand{\thelemma}{C.1}
\begin{lemma}\label{lemma_aux2} Suppose \(\{v_t,\mathcal{H}_t\}_{t \geq 1}\) forms a martingale difference sequences with \(\Ex[v_t^2 \mid \mathcal{H}_{t-1}] = \sigma_v^2 \in (0,\infty)\) and \(a_{t-1}(\theta)\) is adapted to \(\mathcal{H}_{t-1}\) uniformly in \(\theta \inn [\ubtheta,\btheta]\). Then, for any \(m \in \mathbb{N}_+\)
\begin{align}
\normalfont\sumt^n \ssupt (\ma_{t-1}(\theta))^2= \,&  O_p(\ln\,n) \tag{\textit{i}} \\
\normalfont\sumt^n \ssupt (\ma_{t-1}(\theta))^2 v_t^2 = \,&  O_p(\ln\,n) \tag{\textit{ii}} \\
\normalfont\ssupt \bigg\vert\sumt^n \ma_{t-1}(\theta)v_t \bigg\vert = \,&  O_p(\ln\,n). \tag{\textit{iii}}
\end{align}
\end{lemma}

\renewcommand{\thelemma}{C.2}
\begin{lemma}\label{lemma_aux3}
For any \(\theta \in [\ubtheta,\btheta]\), one has \textnormal{($pointwise$)}
\begin{align}
\lni \sumt^n a_t^{\st 2}(\theta) = \,& \sigma_\eps^2 \frac{\theta^2}{2\theta-1}\bigg[1+2\frac{\theta_0(1-c_0)(\theta_0(1+c_0)-1)}{(2c_0\theta_0-1)(c_0\theta_0+\theta-1)}\bigg]+ o_p(1) \tag{\textit{i}} \\
\lni \sumt^n a_t^{\st}a_t^{\st}(\theta) = \,& \sigma_\eps^2 \frac{\theta\theta_0((1 + c_0)\theta_0-1)}{(c_0\theta_0+\theta-1)(2c_0\theta_0-1)}+ o_p(1), \tag{\textit{ii}} 
\end{align}
where \(c_0 = 1-\beta_0\). Moreover,
\begin{align}
\lni \sumt^n \da_t^2 = \,& \sigma_{\da}^2+ o_p(1) \tag{\textit{iii}}\\
\lni \sumt^n a_t^\st \da_t = \,&  \left(\frac{\sigma_a}{\theta_0}\right)^2\frac{\theta_0(1-c_0\theta_0)}{1-\theta_0(1+c_0)} + o_p(1).  \tag{\textit{iv}} 
\end{align}
\end{lemma}

\renewcommand{\thelemma}{C.3}
\begin{lemma}\label{lemma_aux4} For any \(m \in \mathbb{N}_+\) and \(r \in \{2,4\}\)
\[
\Ex^{1/r}\left[(\ma_n(\theta))^r\right] = O(n^{1/2}).
\]
\end{lemma}

\renewcommand{\thelemma}{C.4} 
\begin{lemma}\label{lemma_aux5}  
\(
\Ex\big[\ssupt\ssup\limits_{\mathfrak{a}\inn\mathbb{R}}|Q_n(\theta,\mathfrak{a})-Q_n^\dagger(\theta)|\big] = O(1).
\)
\end{lemma}

\renewcommand{\thelemma}{C.5} 
\begin{lemma}\label{lemma_aux1} 
For any \(m \in \mathbb{N}_+\) and \(r \in \{1,2\}\)
\begin{align}
\normalfont n^r\sumt^n \frac{1}{t^{r-1}}(\mg_{t,n}(\theta))^2 = \;& g(\theta,r,m)+ o(1)\;\; \tag{\textit{i}} \\
n^{1/2}\sumt^n t^{-1/2}\mg_{t,n}(\theta) = \;& \frac{\Gamma(m+1)}{(1/2-\theta)^{m}(2\theta-1)}+ o(1), \tag{\textit{ii}} 
\end{align}
where \(g(\theta,r,m) \coloneqq r^{-2m+1}(2\theta/r-1)^{-2m-1}m(m+(2/r)(\theta/r-1)\theta)\Gamma(2m-1)\). Moreover,
\begin{align}
n\sumt^n g_{t,n}^2(\theta) = \;& \frac{\theta^2}{2\theta-1} + o(1) \tag{\textit{iii}}  \\
n\sumt^n g_{t,n}(\theta)g_{t,n} = \;& \frac{\theta\theta_0}{c_0\theta_0+\theta-1}+ o(1) \tag{\textit{iv}} \\
n\sum_{t \eq 2}^n\sumj^{t-1}g_{t,n}(\theta)g_{j,n}(\theta) g_{j,t}  = \;&  \frac{\theta^2\theta_0}{(2\theta-1)(c_0\theta_0+\theta-1)} + o(1) \tag{\textit{v}} \\
n\sum_{t \eq 2}^n\sumj^{t-1}g_{t,n}(\theta)g_{j,n} g_{j,t}  = \;&  \frac{\theta\theta_0^2}{(2c_0\theta_0-1)(c_0\theta_0+\theta-1)} + o(1) \tag{\textit{vi}} \\
n\sumt^n \dg_{t,n}g_{t,n} = \;& \frac{\theta_0(c_0\theta_0-1)}{(\theta_0(1+c_0)-1)^2}+ o(1) \tag{\textit{vii}} \\
n\sum_{t \eq 2}^n\sumj^{t-1}\dg_{t,n}g_{j,n} g_{j,t}  = \;&  \frac{\theta_0^2(c_0\theta_0-1)}{(2c_0\theta_0-1)(\theta_0(1+c_0)-1)^2} + o(1) \tag{\textit{viii}}\\
n\sum_{t \eq 2}^n\sumj^{t-1}\dg_{t,n}\dg_{j,n} g_{j,t}  = \;&  \frac{\theta_0(\theta_0(2c_0(\theta_0 - 1)\theta_0  + c_0 - \theta_0  + 2) - 1)}{(2\theta_0  - 1)^3((1+c_0)\theta_0 - 1)^2} + o(1). \tag{\textit{ix}}
\end{align}
\end{lemma}

\section[D Proof of Auxiliary Results]{Proof of Auxiliary Results}\label{seq:appC}

The following makes frequently use of the fact that if a sequence of constants \((\varsigma_t)_{t \geq 1}\) is such that \(t\varsigma_t = \varsigma+o(1)\) for some finite constant \(\varsigma\), then \(\lni\sumt^n \varsigma_t = \varsigma + o(1)\).

\subsection{Proof of Lemma \ref{lemma_aux2}}
\underline{Part (\textit{i})-(\textit{ii}):} Recall from the preliminary notes of Appendix \ref{seq:appA} that \((a_n^{(m)}(\theta))^2 = \big[\sumt^n g_{t,n}^{(m)}(\theta)(y_t-\alpha_0)\big]^2,\) where \(y_t -\alpha_0 = \beta_0a_{t-1}^\st+\eps_t\). Note that the claim follows if we can show that \(\Ex[\ssup\limits_{\theta \inn [\ubtheta,\btheta]} (a_n^{(m)}(\theta))^2] = O(n^{-1}).\)  Since \((A+B)^2 \leq 2(A^2+B^2)\) for any \(A,B \in \mathbb{R}\), one gets  
\[
\ssup\limits_{\theta \inn [\ubtheta,\btheta]} (a_n^{(m)}(\theta))^2 \leq 2(A_n^2 + \beta_0^2B_n^2),
\]
where
\[
A_n \coloneqq \ssup\limits_{\theta \inn [\ubtheta,\btheta]}\left\vert\sumt^n\mg_{t,n}(\theta)\eps_t\right\vert,\; B_n \coloneqq \ssup\limits_{\theta \inn [\ubtheta,\btheta]}\left\vert\sumt^n\mg_{t,n}(\theta)a_{t-1}^\st\right\vert.
\]
Similarly, \(A_n^2 \leq 2(A_{1,n}^2+A_{2,n}^2)\), where
\[
A_{1,n} \coloneqq \left\vert\sumt^n\mg_{t,n}(\theta_0)\eps_t\right\vert,\;\; A_{2,n} \coloneqq \ssup\limits_{\theta \inn [\ubtheta,\btheta]}\left\vert\sumt^n(\mg_{t,n}(\theta)-\mg_{t,n}(\theta_0))\eps_t\right\vert.
\]
We have, by Lemma \ref{lemma_aux1} and Assumption \ref{ass:C}, \(\Ex[A_{1,n}^2] = O(n^{-1})\). By the mean-value theorem and the discussion preceding Eq. (3.8) in \cite{lai94}, one obtains the following estimate 
\begin{equation}\label{eq:palma}
\begin{split}
\Ex[A_{2,n}^2] \leq \,&(\btheta-\ubtheta) \Ex\bigg[\int_{[\ubtheta,\btheta]} \bigg[\sumt^n g_{t,n}^{(m+1)}(\theta) \eps_t \bigg]^2 \textsf{d}\theta\bigg]\\
 = \,& O\left(\sumt^n\int_{[\ubtheta,\btheta]} (g_{t,n}^{(m+1)}(\theta))^2\,\textsf{d}\theta\right) = O(n^{-1}).
\end{split}
\end{equation}
The final equality is due Tonelli's theorem in conjunction with the fact that the indefinite integral of the function \(\theta \mapsto g(\theta,r,m)\) defined in item ($i$) of Lemma \ref{lemma_aux1} is non-decreasing in \(\theta\). Turning to \(B_n\), use, again, the bound \(B_n^2 \leq 2(B_{1,n}^2+B_{2,n}^2)\), where
\[
B_{1,n} \coloneqq \left\vert\sumt^n\mg_{t,n}(\theta_0)a_{t-1}^\st\right\vert,\;\; B_{2,n} \coloneqq \ssup\limits_{\theta \inn [\ubtheta,\btheta]}\left\vert\sumt^n(\mg_{t,n}(\theta)-\mg_{t,n}(\theta_0))a_{t-1}^\st\right\vert.
\]
Since, by \citet[Lemma B.4]{may21}, \(\Ex[a_n^{\st 2}] =O(n^{-1})\), it follows readily from Lemma \ref{lemma_aux1} that \(\Ex[B_{1,n}^2] = O(n^{-1})\), while
\begin{equation}
\begin{split}
B_{n,2}^2 \leq \,&  (\btheta-\ubtheta) \int_{[\ubtheta,\btheta]} \left[\sumt^n g_{t,n}^{(m+1)}(\theta) a_{t-1}^\st \right]^2 \textsf{d}\theta \\
\leq \,&  n(\btheta-\ubtheta)  \sumt^n a_{t-1}^{\st 2} \int_{[\ubtheta,\btheta]}(g_{t,n}^{(m+1)}(\theta))^2\textsf{d}\theta, 
\end{split}
\end{equation}
where the mean-value theorem, Cauchy-Schwarz's inequality and Tonelli's theorem have been used. Since \(\Ex[a_{n}^{\st 2}] = O(n^{-1})\), we obtain, by item ($i$) of Lemma \ref{lemma_aux1} and the same argument used in Eq. \eqref{eq:palma}, \(\Ex[B_{n,2}^2] = O(n^{-1})\). This proves part ($i$) and ($ii$).

\noindent \underline{Part ($iii$).} There exist \(\theta_1,\dots,\theta_N\) such that we can cover \([\ubtheta,\btheta] \subset \cup_{j \eq 1}^N \Theta_j\), with the integer \(N \coloneqq N(n)\) satisfying \(N = O(n)\), where \(\Theta_j \coloneqq \{\theta \in (1,\infty): |\theta-\theta_j| \leq 1/n\}\). Therefore, by the triangle inequality, one obtains
\begin{equation}\nn
\begin{split}
\ssup\limits_{\theta \inn [\ubtheta,\btheta]} \frac{1}{\ln\,n}\left\vert \sumt^n \ma_{t-1}(\theta)u_t \right\vert \leq \,& \mmax\limits_{1\lleq j \lleq N}\ssup\limits_{\theta \inn \Theta_j} \frac{1}{\ln\,n}\left\vert \sumt^n(\ma_{t-1}(\theta)-\ma_{t-1}(\theta_j))u_t \right\vert \\
\,& + \mmax\limits_{1\lleq j \lleq N} \frac{1}{\ln\,n}\left\vert \sumt^n\ma_{t-1}(\theta_j)u_t \right\vert.
\end{split}
\end{equation}
Using Cauchy-Schwarz and the mean-value theorem, we can bound the first summand on the right-hand side by
\begin{equation}\nn
\begin{split}
 \mmax\limits_{1\lleq j \lleq N}\ssup\limits_{\theta \inn \Theta_j} |\theta-\theta_j|& \frac{1}{\ln\,n} \left\vert \sumt^n \ssup\limits_{\theta \inn [\ubtheta,\btheta]} a_{t-1}^{(m+1)}(\theta)u_t \right\vert \\
&\;\leq \frac{1}{\sqrt{n\ln\,n}}\sqrt{\frac{1}{\ln\,n}\sumt^n \ssup\limits_{\theta \inn [\ubtheta,\btheta]} (a_{t-1}^{(m+1)}(\theta))^2u_t^2}.
\end{split}
\end{equation}
Thus, the claim follows from
\[
\textnormal{($a$) }\;\;\sumt^n \ssup\limits_{\theta \inn [\ubtheta,\btheta]} (a_{t-1}^{(m+1)}(\theta))^2u_t^2 = O_p(\ln\,n),\;\;\textnormal{($b$) }\;\;\mmax\limits_{1\lleq j \lleq N} \left\vert \sumt^n\ma_{t-1}(\theta_j)u_t \right\vert = O_p(\ln\,n).
\]
Indeed, ($a$) and ($b$) follow directly from part ($ii$) and \citet[Theorem 2.3]{wang21}. This proves part ($iii$).

\subsection{Proof of Lemma \ref{lemma_aux3}}
\underline{Proof of ($i$):} Consider
\begin{equation}
\Ex[a_{t}^{\st 2}(\theta)] = \sumi^tg_{i,t}^2(\theta)\var[y_i] + 2\sum_{i \eq 2}^t\sumj^{i-1}g_{i,t}(\theta)g_{j,t}(\theta)\cov[y_i,y_j]\coloneqq A_t + 2B_t,
\end{equation}
say. Begin with \(A_t\). From \(\cov[a_{t-1}^\st,\eps_t] = 0\), it follows that \(\var[y_t]  = \beta_0^2\var[a_{t-1}^\st] + \var[\eps_1]\), where \(t\,\var[a_{t}^\st] =\sigma_a^2 + o(1)\);
see, e.g. \citet[Lemma B.4]{may21}. Since \(t\sumi^tg_{i,t}(\theta)^2/i = o(1)\) and \(t\sumi^tg_{i,t}^2(\theta) \rightarrow \theta^2/(2\theta-1)\), it follows that \(t\,A_t \rightarrow  \sigma_\eps^2\theta^2/(2\theta-1)\). Turning to the covariances, note that, by Assumption \ref{ass:C},
\(
\cov[y_i,y_j] = \beta_0^2\cov[a_{i-1}^\st,a_{j-1}^\st] + \beta_0\cov[a_{i-1}^\st,\eps_j]
\)
for any \(1 \leq j < i \leq t\), where \(\cov[a_{i}^\st,a_{j}^\st] = \var[a_{j}^\st] \Phi_{i,j+1}\) [see \cite{may21}] and \(\cov[a_{i}^\st,\eps_j] = g_{j,i}\sigma_\eps^2\). Therefore, \(B_t = \textsf{c}\sigma_\eps^2C_t + o(1)\), where
\[
C_t \coloneqq\sum_{i \eq 2}^t\sumj^{i-1}g_{i,t}(\theta)g_{j,t}(\theta) g_{j,i},\quad \textsf{c} \coloneqq \frac{(1-c_0)((1+c_0)\theta_0-1)}{2c_0\theta_0-1}.
\]
It follows from lemma \ref{lemma_aux1} that \(t\,C_t \rightarrow \theta^2\theta_0/((2\theta-1)(c_0\theta_0+\theta-1))\). Upon collecting terms, one obtains thus 
\[
\lni \sumt^n \Ex[a_t^{\st 2}(\theta)] \rightarrow \sigma_\eps^2 \frac{\theta^2}{2\theta-1}\bigg[1+2\frac{\theta_0(1-c_0)(\theta_0(1+c_0)-1)}{(2c_0\theta_0-1)(c_0\theta_0+\theta-1)}\bigg].
\]
The claim follows by Chebychev's inequality because \(\var\big[\sumt^n a_t^{\st 2}(\theta)\big] = o(\ln^{2}n)\). To see this, define, in view of Eq. \eqref{eq:ASTtheta}, \(a_{1,t}^{\st} \coloneqq \sumi^t g_{i,t}(\theta)\eps_i\) and \(a_{2,t}^{\st} \coloneqq \sumi^t g_{i,t}(\theta)a_{i-1}^\st\). We then obtain, 
\[
\var\bigg[\sumt^n a_t^{\st 2}(\theta)\bigg] \leq 8(\var\bigg[\sumt^na_{1,t}^{\st 2}(\theta)\bigg]+\beta_0^4\var[\sumt^na_{2,t}^{\st 2}(\theta)\bigg]).
\]
Now, 
\[
\var\bigg[\sumt^na_{1,t}^{\st 2}(\theta)\bigg] = \sumt^n\var[a_{1,t}^{\st 2}]+2\sumt^{n-1}\sum_{m\eq 1}^{n-t}\cov[a_{1,t}^{\st 2},a_{1,t+m}^{\st 2}].
\]
The first summand on the right-hand side of the preceding display is bounded as \(\var[a_{1,t}^{\st 2}] = O(n^{-2})\). Turning to the covariances, note first that by Assumption \ref{ass:B}, one obtains
\begin{equation}
\begin{split}\nn
\Ex[a_{1,t}^{\st 2}a_{1,t+m}^{\st 2}] 
= \,& \Phi_{t+1,t+m}^2(\theta)\sum_{r,k,i,j = 1}^t g_{i,t}(\theta)g_{j,t}(\theta)g_{r,t}(\theta)g_{k,t}(\theta)\Ex[\eps_i\eps_j\eps_r\eps_k] \\
\,& +\sum_{r,k \eq t+1}^{t+m}\sum_{i,j = 1}^t g_{i,t}(\theta)g_{j,t}(\theta)g_{r,t+m}(\theta)g_{k,t+m}(\theta)\Ex[\eps_i\eps_j\eps_r\eps_k], 
\end{split}
\end{equation}
where it has been used that \(g_{i,t+m}(\theta) = \Phi_{t+1,t+m}(\theta)g_{i,t}(\theta)\). Hence, using again Assumption \ref{ass:B}, and \(\Ex[a_{t,1}^{\st 2}(\theta)] = \sigma_\eps^2\sumi^tg_{i,t}^2(\theta)\), one gets
\begin{equation}
\begin{split}\nn
\cov[a_{1,t}^{\st 2}a_{1,t+m}^{\st 2}] = \,& \Phi_{t+1,t+m}^2(\Ex[\eps_1^4]\sumi^tg_{i,t}^4(\theta)+6\sigma_\eps^4\sum_{i \eq 2}^t\sumj^{i-1}g_{i,t}^2(\theta)g_{j,t}^2(\theta))\\
\,&-\sigma_\eps^4\Phi_{t,t+m}^2(\theta)(\sumi^tg_{i,t}^2(\theta))^2 = O(g_{t,t+m}^2(\theta)),
\end{split}
\end{equation}
 where the last equality follows from Lemma \ref{lemma_aux1}. Now, note that 
\[
\sumt^{n-1}\sum_{m\eq 1}^{n-t}g_{t,t+m}^2(\theta) = \sum_{t \eq 2}^{n}\sum_{m\eq 1}^{t-1}g_{m,t}^2(\theta) = \sumt^nO(t^{-1}) = O(\ln\,n),
\]
where the penultimate equality is due to Lemma \ref{lemma_aux1}. This shows that \(\var\big[\sumt^na_{1,t}^{\st 2}(\theta)\big] = o(\ln\,n)\). Similarly, using \(\var[a_n^\st] = O(n^{-1})\), we obtain \(\var\big[\sumt^na_{2,t}^{\st 2}(\theta)\big] = o(\ln\,n)\), which completes the proof. \underline{Proof of ($ii$):}  Use that \(\Ex[a_j^\st\eps_i] = 1\{j \geq i\}\sigma_\eps^2 g_{i,j}\) to obtain
\[
\Ex[a_{t}^\st a_{t}^\st(\theta)] = \sigma_\eps^2\left[\sumj^t g_{j,t}(\theta)g_{j,t}+ \beta_0\sum_{i \eq 2}^t\sumj^{i-1}g_{i,t}(\theta)g_{j,t} g_{j,i}\right].
\]
Since, by Lemma \ref{lemma_aux1}, 
\[
t\sumj^t g_{j,t}(\theta)g_{j,t} \rightarrow \theta\theta_0/(c_0\theta_0+\theta-1),
\] it follows from \(t\,\sum_{i \eq 2}^t\sumj^{i-1}g_{i,t}(\theta)g_{j,t} g_{j,i} \rightarrow \theta\theta_0^2/((2c_0\theta_0-1)(c_0\theta_0+\theta-1))\), that 
\[
\lni \sumt^n \Ex[a_t^\st a_t^{\st 2}(\theta)] \rightarrow \sigma_\eps^2 \frac{\theta\theta_0((1 + c_0)\theta_0-1)}{(c_0\theta_0+\theta-1)(2c_0\theta_0-1)}.
\]
Similar to part ($i$), the claim follows by Chebychev's inequality because \(\var\big[\sumt^n a_t^{\st}a_t^{\st}(\theta)\big] = o(\ln^{2}n)\). \underline{Proof of ($iii$):} Consider
\begin{equation} 
\Ex[\da_t^2] = \var[\da_t]= \sumj^t  \dg_{j,t}^2\var[y_k] + 2\sum_{i \eq 2}^t\sumj^{i-1}\dg_{i,t}\dg_{j,t}\cov[y_k,y_j] \coloneqq A_t + 2B_t,
\end{equation}
say. We obtain \(tA_t = (1+2\theta_0(\theta_0-1))/(2\theta_0-1)^3 + o(1)\) from the same arguments used in the proof of part ($i$) together with Lemma \ref{lemma_aux1}. Similar to the proof of part ($i$), we also obtain \(B_t = \var[\eps_1]\textsf{c}C_t + o(1)\), where
\[
C_t \coloneqq\sum_{k \eq 2}^t\sum_{j \eq 1}^{k-1}\dg_{k,t}\dg_{j,t} g_{j,k},
\]
and \(\textsf{c}\) has been defined above. Use of item ($viii$) of Lemma \ref{lemma_aux1} and collecting terms yields \(\lni\sumt^n\Ex[\da_t^2] = \sigma_{\da}^2 + o(1)\). Similar to part ($i$), the claim follows by Chebychev's inequality. \underline{Proof of ($iv$):} Similar to the proof of part ($ii$), we obtain
\[
\Ex[a_{t}^\st \da_{t}] = \sigma_\eps^2\left[\sumj^t \dg_{j,t}g_{j,t}+ \beta_0\sum_{i \eq 2}^t\sumj^{i-1}\dg_{i,t}g_{j,t} g_{j,i}\right]+o(1).
\]
As above, the claim follows from Lemma \ref{lemma_aux1} and, similar to part ($i$), from Chebychev's inequality.

\subsection{Proof of Lemma \ref{lemma_aux4}}
Let \(\ma_{1,n} \coloneqq \sumt^n\mg_{t,n}(\theta_0)\eps_t\) and \(\ma_{2,n} \coloneqq \sumt^n\mg_{t,n}a_{t-1}^\st\). For \(r\in \{2,4\}\), we thus obtain from the \(c_r\)-inequality
\[
\Ex[(\ma_n(\theta_0))^r] \leq 2^{r-1}(\Ex[(\ma_{1,n}(\theta_0))^r]+\beta_0^r\Ex[(\ma_{2,n}(\theta_0))^r]).
\]
\underline{Case \(r = 2\).} The claim follows directly from the proof of item (\textit{i}) of Lemma \ref{lemma_aux3}. \underline{Case \(r = 4\).} By Assumption \ref{ass:C}, \(\Ex[\eps_{i}\eps_{j}\eps_{r}\eps_{s}] \neq 0\) if either all indices are equal, or if two indices are equal within pairs but unequal across pairs (3 occurences). Thus, by lemma item (\textit{i}) of \ref{lemma_aux1}.
\begin{equation}
\begin{split}
\Ex[(\ma_{1,n}(\theta_0))^4] = \,& \Ex[\eps_1^4]\sumt^n(\mg_{t,n}(\theta_0))^4+\sigma_\eps^46 \sum_{t \eq 2}^n\sumi^{t-1}(\mg_{t,n}(\theta_0)\mg_{i,n}(\theta_0))^2\\
 \leq \,& 3\Ex[\eps_1^4]\bigg[\sumt^n(\mg_{t,n}(\theta_0))^2\bigg]^2 = O(n^{-2}). 
\end{split}
\end{equation}
Similarly, since \(\Ex[a_n^{\st 4}] \leq \textsf{C}n^{-2}\) [see \citet[Lemma B.4]{may21}] for some \(\textsf{C} \in (0,\infty)\) repeated application of Cauchy-Schwarz's inequality yields
\begin{equation}
\Ex[(\ma_{2,n}(\theta_0))^4] \leq \textsf{C}\bigg[\sumt^nt^{-1/2}\mg_{t,n}(\theta_0)\bigg]^4 = O(n^{-2}),
\end{equation}
 using item (\textit{ii}) of Lemma \ref{lemma_aux1}. This proves the claim.

\subsection{Proof of Lemma \ref{lemma_aux5}}
Let \(Q_n(\theta) \coloneqq Q_n(\theta,\mathfrak{a}_0)\) and note that, by the triangle inequality and Eq. \eqref{eq:AST}, 
\begin{equation}
\begin{split}\nn
|Q_n(\theta,\mathfrak{a})-Q_n(\theta)| \leq \,&\sumt^n(a_{t-1}(\theta)-a_{t-1}(\theta,\mathfrak{a}))^2+\sumt^n|a_{t-1}(\theta)-a_{t-1}(\theta,\mathfrak{a})||u_t|\\
=\,&(\mathfrak{a}-\mathfrak{a}_0)^2\sumt^n \Phi_{t-1}^2(\theta)+|\mathfrak{a}-\mathfrak{a}_0|\sumt^n|\Phi_{t-1}(\theta)||u_t|
\end{split}
\end{equation}
Since \(\Phi_n(\theta) =O(t^{-\theta})\) and \(\ubtheta > 1\), we obtain \(\Ex[\ssupt\ssup\limits_{\mathfrak{a}\inn\mathbb{R}}|Q_n(\theta,\mathfrak{a})-Q_n(\theta)|] = O(1)\). Taking into account
\begin{equation}\nn
|Q_n(\theta)-Q_n^\dagger(\theta)| \leq (\mathfrak{a}_0-\alpha_0)^2\sumt^n \Phi_{t-1}^2(\theta)+|\mathfrak{a}_0-\alpha_0|\sumt^n|\Phi_{t-1}(\theta)||u_t|,
\end{equation}
similar arguments as above yield \(\Ex[\ssupt|Q_n(\theta)-Q_n^\dagger(\theta)|] = O(1)\). The claim then follows by the triangle inequality.

\subsection{Proof of Lemma \ref{lemma_aux1}}
\textbf{Part (\textit{i}).} Suppose, \textit{w.l.o.g.}, that if \(\theta \in \mathbb{N}_+\), then \(\theta \leq t \leq n\). Note that if this does not hold, then \(g_{t,n}(\theta) = 0\) so that all derivatives are zero. It is instructive to consider first the first two derivatives of \(g_{t,n}(\theta)\); i.e., \(m \in \{1,2\}\).  By the product rule, one obtains 
\[
\frac{\textsf{d}\,g_{t,n}(\theta)}{\textsf{d}\,\theta} \eqqcolon \dg_{t,n}(\theta) = g_{t,n}(\theta)\bigg[\sum_{i \eq t+1}^n(\theta-i)^{-1}+1/\theta\bigg] 
\]
and
\[
\frac{\textsf{d}^2\,g_{t,n}(\theta)}{\textsf{d}\,\theta^2} \eqqcolon \ddg_{t,n}(\theta) = g_{t,n}(\theta)\bigg[\bigg(\sum_{i \eq t+1}^n(\theta-i)^{-1}+1/\theta\bigg)^2-\sum_{i \eq t+1}^n(\theta-i)^{-2}-1/\theta^2\bigg].
\]
The preceding displays can be expressed in terms of the gamma function \(\Gamma(z)\) and its  \((n+1)\)-th logarithmic derivatives, the polygamma functions \(\psi^{(n)}(z)\), which are defined for any \(n \in \mathbb{N}\) and all \(z \in \mathbb{R}\setminus\{-1,-2,\dots\}\):
\begin{align}
\dg_{t,n}(\theta) =  \,& g_{t,n}(\theta)\left[\psi(j+1-\theta)-\psi(t+1-\theta)+1/\theta\right] \nn\\
\ddg_{t,n}(\theta) =\,& g_{t,n}(\theta)\bigg\{2\big[\big(\psi(j+1-\theta)-\psi(t+1-\theta)\big)/\theta-\psi(j+1-\theta)\psi(t+1-\theta)\big]\nn \\
\,& \,\hspace*{.45cm} - \big(\psi^{(1)}(j+1-\theta)-\psi^{(1)}(t+1-\theta)\big) +\psi(j+1-\theta)^2+\psi(t+1-\theta)^2\bigg\}, \nn
\end{align}
while
\[
g_{t,n}(\theta) = \frac{\theta}{t} \frac{\Gamma(t+1)}{\Gamma(n+1)}\frac{\Gamma(n+1-\theta)}{\Gamma(t+1-\theta)},
\]
where we use for the digamma function \(\psi(z) \coloneqq \psi^{(0)}(z)\). Next, use that \(\Gamma(z+\alpha)/\Gamma(z+\beta) = z^{\alpha-\beta}(1+O(z^{-1}))\), \(\psi(z) = \ln\,z+O(z^{-1})\), and \(\psi^{(n)}(z) = O(z^{-n})\) for for all \(\alpha,\beta \in \mathbb{R}\) and large \(z>0\); see, e.g., \cite{te51}. Thus, it is not difficult to show that
\[
n^{r}\sumt^nt^{r-1}(\dg_{t,n}^2(\theta)-\df_{t,n}^2(\theta)) = o(1),\;\; n^{r}\sumt^nt^{r-1}(\ddg_{t,n}^2(\theta)-\ddf_{t,n}^2(\theta)) = o(1),\;\; r \in \{1,2\},
\]
where \(\df_{t,n}(\theta) =  f_{t,n}(\theta)(\ln(t/n)+1/\theta), \ddf_{t,n}(\theta)=f_{t,n}(\theta)\ln(t/n)(\ln(t/n)+2/\theta)\) denote the two first derivatives of \(f_{t,n}(\theta) \coloneqq \theta t^{\theta-1}n^{-\theta}\) with respect to \(\theta\). More generally, it can be shown that
\[
n^{r}\sumt^nt^{r-1}\bigg[\bigg(\frac{\textsf{d}^m\,g_{t,n}(\theta)}{\textsf{d}\,\theta^m}\bigg)^2 -\bigg(\frac{\textsf{d}^m\,f_{t,n}(\theta)}{\textsf{d}\,\theta^m}\bigg)^2\bigg] = o(1),\;\; r \in \{1,2\},
\]
where
\[
\frac{\textsf{d}^m\,f_t(\theta)}{\textsf{d}\,\theta^m} =  f_{t,n}(\theta)\ln^{m-1}(t/n)(\ln(t/n)+m/\theta),\;\;m \in \mathbb{N}_+.
\]
Some calculations reveal
\[
n^r\int_1^n\frac{1}{t^{r-1}} \bigg[ \frac{\textsf{d}^m\,f_{t,n}(\theta)}{\textsf{d}\,\theta^m} \bigg]^2  \textsf{d}t \rightarrow  r^{-2m+1}(2\theta/r-1)^{-2m-1}m(m+(2/r)(\theta/r-1)\theta)\Gamma(2m-1),
\]
while by the Euler-Maclaurin formula
\[
n^r\bigg(\sumt^n \frac{1}{t^{r-1}}\bigg[ \frac{\textsf{d}^m\,f_{t,n}(\theta)}{\textsf{d}\,\theta^m} \bigg]^2 -\int_1^n\frac{1}{t^{r-1}} \bigg[ \frac{\textsf{d}^m\,f_{t,n}(\theta)}{\textsf{d}\,\theta^m} \bigg]^2\textsf{d}t\bigg) = o(1) 
\]
for any \(m \in \mathbb{N}_+\) and \(r\in\{1,2\}\) as \(n \rightarrow \infty\). \textbf{Part (\textit{ii}).}  Upon noting that \[n^{1/2}\int_1^n t^{-1/2} f_{t,n}(\theta)\textsf{d}t \rightarrow  \Gamma(m+1)/((1/2-\theta)^{m}(2\theta-1)),\] the claim follows by similar arguments used in part (\textit{i}). \textbf{Part (\textit{iii}).} Upon noting that \[n\int_1^n f_{t,n}^2(\theta)\textsf{d}t \rightarrow \theta^2/(2\theta-1),\] the claim follows by similar arguments used in part (\textit{i}). \textbf{Parts (\textit{iv})-(\textit{ix}).} Follows by similar arguments.

		\end{document}